\documentclass[pre,onecolumn,showpacs]{revtex4}

\usepackage{amssymb}
\usepackage{graphicx}
\usepackage{colordvi}
\usepackage{color}
\usepackage{subfigure}

\begin{document}

\title{Anomalous currents in driven XXZ chain with boundary twisting at weak coupling
or weak driving }
\author{Vladislav Popkov$^{1,2}$ and Mario Salerno$^{3}$}

\begin{abstract}
The spin $1/2$ $XXZ$ chain driven out of equilibrium by coupling
with boundary reservoirs targeting perpendicular spin orientations
in $XY$ plane, is investigated. The existence of an anomaly in the
nonequilibrium steady state (NESS) at the isotropic point
$\Delta=1$ is demonstrated both in the weak coupling and weak
driving limits. The nature of the anomaly is studied analytically
by calculating exact NESS for small system sizes, and
investigating steady currents. The spin current at the points
$\Delta=\pm1$ has a singularity which leads to a current
discontinuity when either driving or coupling vanish, and the
current of energy develops a twin peak anomaly. The character of
singularity is shown to depend qualitatively on whether the system
size is even or odd.
\end{abstract}

\maketitle

\address{$^{1}$ Dipartimento di Fisica e Astronomia, Universit\`a di Firenze, via G.
Sansone 1, 50019 Sesto Fiorentino, Italy}
\address{$^{2}$ Max Planck Institute for Complex Systems, N\"othnitzer Stra{\ss }e 38,
01187 Dresden, Germany}
\address{$^{3}$ Dipartimento di Fisica ``E.R. Caianiello'', Universit\`a di Salerno,
via ponte don Melillo, 84084 Fisciano (SA), Italy}

\section{Introduction}

The coupling of a quantum system with an environment or under a continuous
measurement is often modelled via a quantum Master equation in the Lindblad
form \cite{Petruccione,PlenioJumps}. Unlike a quantum system evolving
coherently, the time evolution of which depends on the initial state, a
quantum system with pumping tends to a nonequilibrium steady state,
independent of the initial conditions. Recently there has been a large
interest in investigating chains of quantum spins driven towards
nonequilibrium steady states by applying gradients (pumping) at the edges. An
important role in these studies is played by the one dimensional driven XXZ
chain. The equilibrium XXZ model remains a subject of intensive investigations
via various methods
\cite{reviewBrenig07,zotos,KlumperLectNotes2004,Bosonisation,SchollwoeckRev05}%
. In experiments, many transport characteristics in quasi one-dimensional spin
chain materials can be measured \cite{HessBallistic2010}. In the context of
dissipative dynamics, recent numerical and analytic studies of a driven XXZ
chain show a number of remarkable features like long range order and quantum
phase transitions far from equilibrium \cite{TP_Pizorn08}, negative
differential conductivity of spin and energy
\cite{BenentiPRB2009,ZunkovichJStat2010ExactXY}, solvability
\cite{Pros08,ProsenExact2011,MPA} etc..

Recently it was shown that an isotropic point $\Delta=1$, a weakly driven
$XXZ$ spin chain has an anomalous spin transport \cite{znidaricprl2011}. The
current-carrying nonequilibrium steady state was obtained by bringing an $XXZ$
spin chain in contact with reservoirs aligning the boundary spins along the
$Z$-axis. The boundary gradient was small, which allows to think in terms of
linear response of a system to the weak perturbation along the anisotropy axis.

From a standard theory of linear response \cite{KuboBook} one expects that the
result of linear response will not depend on a form of boundary perturbation
as long as it remains sufficiently weak. Respectively, an anomaly seen with
one type of perturbation, should be visible with another type of perturbation
as well. On the other hand, the linear response theory is valid for system in
thermodynamic equilibrium at a certain temperature, while the driven chains
with dissipative boundary driving are intrinsically nonequilibrium systems. At
"zero gradient" the NESS describes a system without currents, in a maximally
mixed quantum state, which corresponds to a Gibbs ensemble with an infinite
temperature. To test further the linear response regime in a situation of a
nonequilibrium, we investigate a weakly driven $XXZ$ chain, where the weak
boundary gradient at the boundaries is applied in the $XY$ direction,
transverse to the anisotropy axis \cite{Lindblad2011,PopkovXYtwist}. Note that
there are two complementary ways of realizing a weak drive in a Master
equation formalism: (i) to set a weak boundary gradient, but finite or strong
coupling to boundary reservoirs (ii) to set a weak coupling to reservoir, but
finite or strong boundary gradient.

Interestingly, in both contexts (either weak gradient $\kappa\ll1$ or weak
coupling $\Gamma\ll1$) we see a distinct anomaly in the NESS. The anomaly
becomes a singularity as $\Gamma\rightarrow0$ or $\kappa\rightarrow0$.  of the
magnetization current in the isotropic point $\Delta=1$. For weak coupling the
anomaly sets in for $N\geq2$ and for weak driving for $N\geq4$. The presence
of an anomaly is seen in practically all observables. In present communication
we focus on singular behaviour of the steady currents of spin $j$ and energy
$J^{E}$, and one-point correlations at the boundaries. The full analytic
treatment is reported for spin current.  In the limit $\kappa\rightarrow0$ or
$\Gamma\rightarrow0$ we observe a non-analyticity in the dependence
$j(\Delta)$: the spin current is an analytic smooth function of $\Delta$
everywhere except at isotropic point $\Delta=1$ where its value is different.
The nature of singularity, in addition, depends on the parity of system size,
namely on whether the system size is an even or an odd number.

Other observables e.g. current of energy, density profiles etc. also have a
singular behaviour at $\Delta=\pm1$. Thus, we can complement the result of
Znidaric \cite{znidaricprl2011} with prediction of further anomalies at the
isotropic point. In particular, we demonstrate that the energy current also
develops an anomaly at the isotropic point, but with a different scaling and
of a qualitatively different form.

The plan of the paper is the  following. We introduce the model in the
Sec.\ref{sec::The model}. The cases of weak coupling and weak driving are
treated separately in Sec.\ref{sec::WeakCoupling} and
Sec.\ref{sec::WeakGradient}. In the conclusion we summarize our findings.

\section{The model}

\label{sec::The model}

We study an open chain of $N$ quantum spins in contact with boundary
reservoirs. The time evolution of the reduced density matrix $\rho$ is
described by a quantum Master equation in the Lindblad form \cite{Petruccione}%
, \cite{PlenioJumps} (here and below we set $\hbar=1$)%
\begin{equation}
\frac{\partial\rho}{\partial t}=-i\left[  H,\rho\right]  +\Gamma
_{L}\mathcal{L}_{L}[\rho]+\Gamma_{R}\mathcal{L}_{R}[\rho],\label{LME}%
\end{equation}
where $H$ is the Hamiltonian of an open $XXZ$ spin chain with an anisotropy
$\Delta$
\begin{equation}
H=\sum_{k=1}^{N-1}\left(  \sigma_{k}^{x}\sigma_{k+1}^{x}+\sigma_{k}^{y}%
\sigma_{k+1}^{y}+\Delta\sigma_{k}^{z}\sigma_{k+1}^{z}\right)
,\label{Hamiltonian}%
\end{equation}
while $\mathcal{L}_{L}$ and $\mathcal{L}_{R}$ are Lindblad dissipators
favoring a relaxation of boundary spins $k=1$ and $k=N$ towards states
described by  density matrices acting on a single site  $\rho_{L}$ and
$\rho_{R}$, i.e. $\mathcal{L}_{L}[\rho_{L}]=0$ and $\mathcal{L}_{R}[\rho
_{R}]=0$. The Lindblad actions $\mathcal{L}_{L}$ and $\mathcal{L}_{R}$ are
chosen in the form
\begin{eqnarray*}
\mathcal{L}_{L}[\rho]  & =-\frac{1}{2}\sum_{m=1}^{2}\left\{
\rho,W_{m}^{\dag
}W_{m}\right\}  +\sum_{m=1}^{2}W_{m}\rho W_{m}^{\dag},\label{LindbladAction}\\
\mathcal{L}_{R}[\rho]  & =-\frac{1}{2}\sum_{m=1}^{2}\left\{  \rho,V_{m}^{\dag
}V_{m}\right\}  +\sum_{m=1}^{2}V_{m}\rho V_{m}^{\dag},\nonumber
\end{eqnarray*}
where $W_{m}$ and $V_{m}$ are polarization targeting Lindblad
operators, which act on the first and on the last spin
respectively, $W_{1}=\sqrt
{\frac{1-\kappa}{2}}(\sigma_{1}^{z}+i\sigma_{1}^{x})$,
$W_{2}=\sqrt
{\frac{1+\kappa}{2}}(\sigma_{1}^{z}-i\sigma_{1}^{x})$,
$V_{1}=\sqrt
{\frac{1+\kappa}{2}}(\sigma_{N}^{y}+i\sigma_{N}^{z})$,
$V_{2}=\sqrt
{\frac{1-\kappa}{2}}(\sigma_{N}^{y}-i\sigma_{N}^{z})$. In absence
of the unitary term in (\ref{LME}) the boundary spins relax with a
characteristic time $\Gamma_{L}^{-1},\Gamma_{R}^{-1}$
\cite{BoundaryRelaxationTimes} to specific states described via
the one-site density matrices $\rho_{L}$ and
$\rho_{R}$, satisfying $\mathcal{L}_{L}[\rho_{L}]=0$ and $\mathcal{L}_{R}%
[\rho_{R}]=0$, where
\begin{eqnarray*}
\rho_{L}  & =\frac{1}{2}\left(  I-\kappa\sigma_{1}^{y}\right)  \label{RoL}\\
\rho_{R}  & =\frac{1}{2}\left(  I+\kappa\sigma_{N}^{x}\right)  .\label{RoR}%
\end{eqnarray*}
From the definition of a one-site density matrix $\rho_{one-site}=\frac{1}%
{2}\left(  I+{\textstyle\sum}\langle\sigma^{\alpha}\rangle\sigma^{\alpha
}\right)  $, we see that the Lindblad superoperators $\mathcal{L}_{L}$ and
$\mathcal{L}_{R}$ indeed try to impose a twisting angle of $\pi/2$ in $XY$
plane between the first and the last spin \cite{TwistingRemark}. The twisting
gradient drives the system in a steady-state with currents. In the following
we restrict to a symmetric choice
\begin{equation}
\Gamma_{L}=\Gamma_{R}=\Gamma\label{SymmetryU-Uprime}%
\end{equation}
for spin current while $\Gamma_{L}\neq\Gamma_{R}$ is chosen for studying the
energy current\cite{energyRemark}. The parameter $\kappa$ determines the
amplitude of the gradient between the left and right boundary, and therefore
plays a fundamentally different role from the coupling strength $\Gamma$. The
limits $\kappa=1$ and $\kappa\ll1$ will be referred to as the strong driving
and weak driving case respectively. The two limits describe different physical
situations as exemplified in other NESS\ studies. For symmetric choice
(\ref{SymmetryU-Uprime}) the steady state has a global symmetry
\cite{PopkovXYtwist},\cite{SP2012}
\begin{equation}
\rho(N,\Delta)=\Sigma_{x}U_{rot}R\rho(N,\Delta)RU_{rot}^{+}\Sigma
_{x},\label{SymmetryGlobal}%
\end{equation}
where $R(A\otimes B\otimes...\otimes C)=(C\otimes....\otimes B\otimes A)R$ is
a left-right reflection, and the diagonal matrix $U_{rot}=diag(1,i)^{\otimes
_{N}}$ is a rotation in $XY$ plane, $U_{rot}\sigma_{n}^{x}U_{rot}^{+}%
=\sigma_{n}^{y}$, $U_{rot}\sigma_{n}^{y}U_{rot}^{+}=-\sigma_{n}^{x}$, and
$\Sigma_{x}=(\sigma^{x})^{\otimes_{N}}$. In addition, there are further
transformations, see \cite{PopkovXYtwist} for details, relating the NESS\ for
positive and negative $\Delta$,
\begin{equation}
\rho(2N,-\Delta)=U\rho^{\ast}(2N,\Delta)U\label{SymmetryEven}%
\end{equation}%
\begin{equation}
\rho(2N+1,-\Delta)=\Sigma_{y}U\rho^{\ast}(2N+1,\Delta)U\Sigma_{y}%
,\label{SymmetryOdd}%
\end{equation}
where $\Sigma_{y}=(\sigma^{y})^{\otimes_{N}},$ and $U=%
{\textstyle\prod\limits_{n\,odd}}
\otimes\sigma_{n}^{z}$, and conjugate is taken in the basis where $\sigma^{z}$
is diagonal. In view of the above symmetries, below we restrict to the case of
positive $\Delta\geq0$.

We shall consider two particular limits:

A) weak coupling, $\Gamma\ll1$, strong driving (large gradient) $\kappa=1$).

B) weak gradient $\kappa\ll1$, strong coupling $\Gamma^{-1}\ll1$

Both limits show qualitatively similar singular behaviour as $\kappa
\rightarrow0$ or $\Gamma\rightarrow0$, at the isotropic point $\Delta=1$ as
demonstrated below.

\section{Case of strong driving $\kappa=1$ and weak coupling $\Gamma\ll1$}

\label{sec::WeakCoupling} To investigate the system in the case of large
boundary gradients $\kappa=1$ and weak coupling $\Gamma\ll1$, we solve the
linear system of equations which determine the full steady state,
analytically, making use of the global symmetry (\ref{SymmetryGlobal})
\cite{SP2012} which decreases the number of unknown variables by roughly a
factor of $2$. For $N=2,3,4$ we have indeed $9$, $31$ and $135$ real unknowns,
respectively, instead of $2^{2N}-1$ real unknowns for the full Hilbert space.
For $N=2$ the current $j(\Gamma,\Delta)$ is readily obtained as
\begin{equation}
j(\Gamma,\Delta)=\frac{8 \Gamma^{4} (3 + 6 \Gamma^{2} + 2 \Delta^{4})}{24
\Gamma^{6} + 3 (\Delta^{2} -1)^{2} + 2 \Gamma^{4} (7 \Delta^{4}+30) +
\Gamma^{2} (2 \Delta^{4} + 13 \Delta^{2} + 30)},
\end{equation}
from which we see that at the points $\Delta=\pm1$ and in the small coupling
limit, it scales quadratically with $\Gamma$,
\begin{equation}
\lim_{\Gamma\rightarrow0}\lim_{\Delta\rightarrow1}\left.  \frac{j(\Gamma
)}{\Gamma^{2}}\right\vert _{N=2}=\frac{8}{9}. \label{J2singularity}%
\end{equation}
Note, however, that for $\Delta\neq\pm1$ the current scales as $\Gamma^{4}$,
i.e
\begin{equation}
\lim_{\Gamma\rightarrow0}\left.  \frac{j(\Gamma)}{\Gamma^{4}} \right\vert
_{N=2}= \frac{8 (2 \Delta^{2} + 3)}{3 (\Delta^{2} -1)^{2}}
\label{N=2singularity1}%
\end{equation}

For $N=3$ the current $j(\Gamma,\Delta)$ at $\Delta=1$, still scales as
$\Gamma^{2}$,
\begin{equation}
\lim_{\Gamma\rightarrow0}\lim_{\Delta\rightarrow1}\left.  \frac{j(\Gamma
)}{\Gamma^{2}}\right\vert _{N=3}=\frac{16}{9} \label{N=3singularity0}%
\end{equation}
but a different scaling behavior, with respect to $N=2$ case, is obtained for
other $\Delta$ points, where the $j$ is found to scale also quadratically,
\begin{equation}
\lim_{\Gamma\rightarrow0}\left.  \frac{j(\Gamma)}{\Gamma^{2}}\right\vert
_{N=3}=\frac{6\Delta\left(  9\Delta^{2}+59\right)  }{9\Delta^{4}+155\Delta
^{2}+416},
\end{equation}
Taking the limit $\Delta\rightarrow1$ in the above, we obtain
\begin{equation}
\lim_{\Delta\rightarrow1}\lim_{\Gamma\rightarrow0}\left.  \frac{j(\Gamma
)}{\Gamma^{2}}\right\vert _{N=3}=\frac{102}{145}, \label{N=3singularity1}%
\end{equation}
different from (\ref{N=3singularity0}). Consequently, the limits
$\Delta\rightarrow1$ and $\Gamma\rightarrow0$ do not commute.  In more
details, we have
\begin{equation}
\lim_{\Delta\rightarrow1}\left.  \frac{j(\Gamma)}{\Gamma^{2}}\right\vert
_{N=3}= \frac{16 (3+\Gamma^{2})} {27+52\Gamma^{2}+12\Gamma^{4}},
\end{equation}
from which Eq(\ref{N=3singularity0}) is straightforwardly obtained.
Alternatively, one can see the presence of a singularity at all even orders of
expansion of $j(\Gamma)$: starting from the order $4$ of the expansion, all
even terms $\left.  (\partial^{m}j_{0}(\Gamma,\Delta)/\partial\Gamma
^{m})\right\vert _{\Gamma=0}$ contain poles at $\Delta=\pm1$ of the type
$1/(\Delta^{2}-1)^{m-2}$, e.g.
\begin{equation}
\frac{\partial^{m}j_{0}(\Gamma,\Delta)}{\partial\Gamma^{m}} \vert_{\Gamma=0}=
\frac{(-1)^{\frac m2 +1}}{(\Delta^{2}-1)^{m-2}} Q(\Delta)
\end{equation}
with $Q(\Delta)$ a rational function which is regular at points $\Delta=\pm1$.

For $N=4$ one can see that the current at $\Delta=\pm1$ scales with $\Gamma$
as:
\begin{equation}
\lim_{\Delta\rightarrow\pm1}\left.  \frac{j(\Gamma)}{\Gamma^{2}}\right\vert
_{N=4}= \frac{8(27+ 46 \Gamma^{2} +10 \Gamma^{4})}{ 3(27 + 135 \Gamma^{2} +
144 \Gamma^{4} +28 \Gamma^{6})}%
\end{equation}
from which it follows that in the small coupling limit the current reduces to
\begin{equation}
\lim_{\Gamma\rightarrow0}\lim_{\Delta\rightarrow\pm1}\left.  \frac{j(\Gamma
)}{\Gamma^{2}}\right\vert _{N=4}= \frac{8}{3}.\label{J4singularity}%
\end{equation}
At all other points $\Delta\neq\pm1$, however, the current scales as for the
$N=2$ case, e.g
\begin{equation}
\lim_{\Gamma\rightarrow0}\left.  \frac{j(\Gamma)}{\Gamma^{4}}\right\vert
_{N=4}=O\left(  \frac{1}{(\Delta^{2} -1)^{2}}\right) .
\label{JevenSingularity0}%
\end{equation}

\begin{figure}[ptbh]
\begin{center}
\subfigure[\label{fig:XYaa}]
{\includegraphics[width=0.4\textwidth]{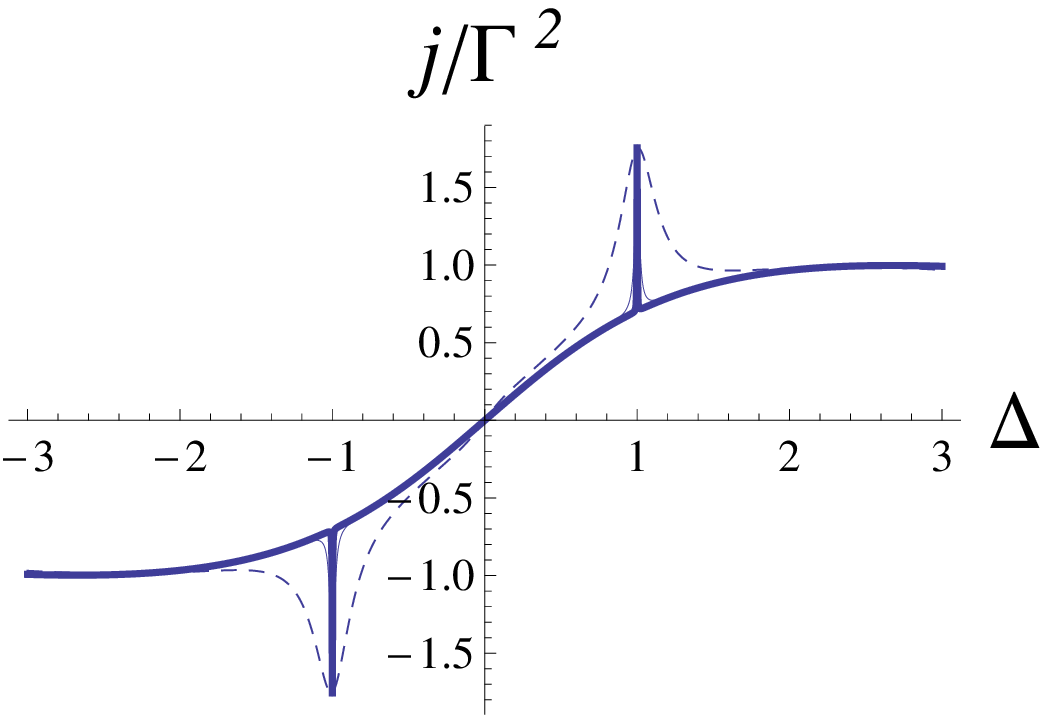}} \qquad
\subfigure[\label{fig:XYbb}]
{\includegraphics[width=0.4\textwidth]{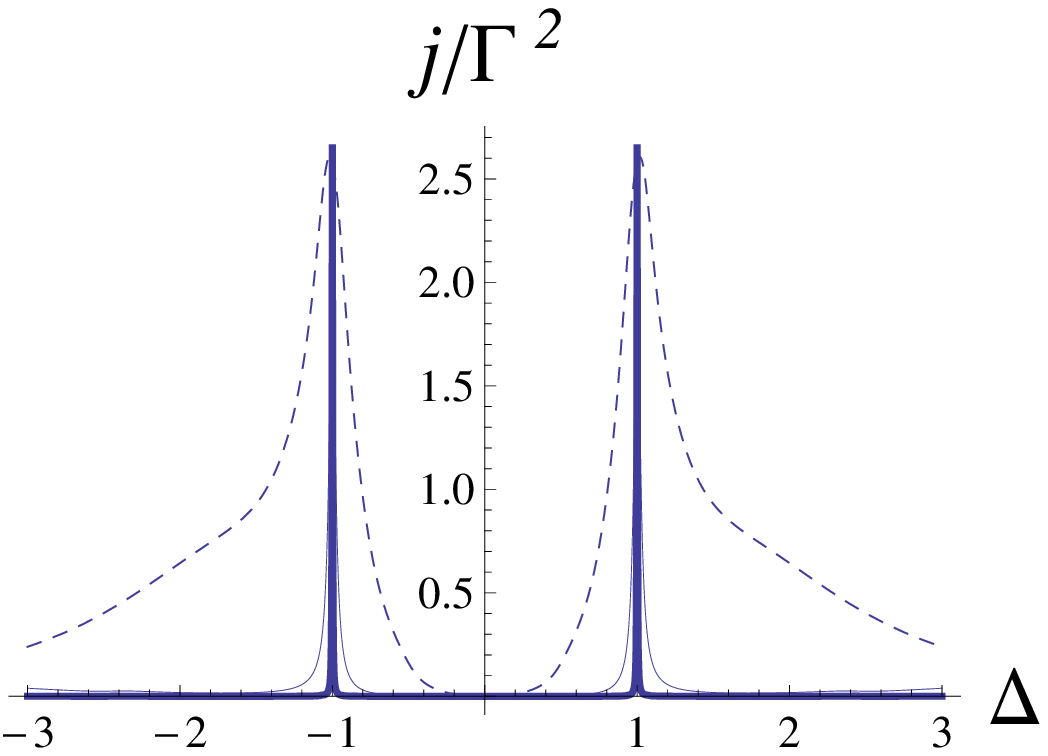}} \qquad
\subfigure[\label{fig:XYc}]{\includegraphics[width=0.4\textwidth]{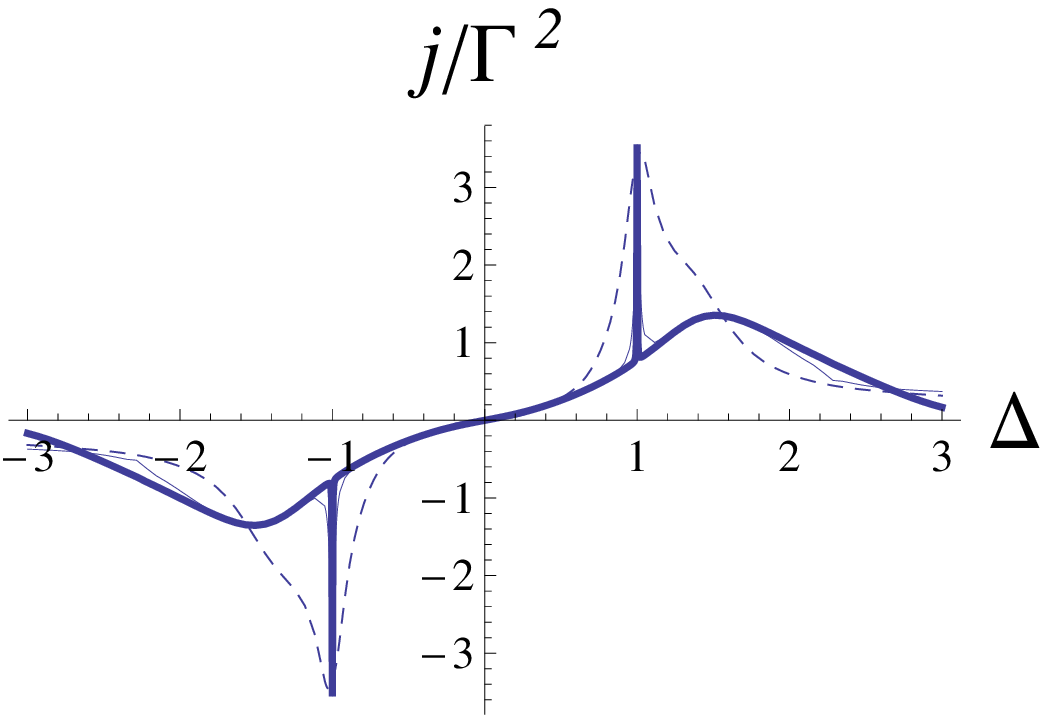}}
\qquad
\subfigure[\label{fig:XYd}]{\includegraphics[width=0.4\textwidth]{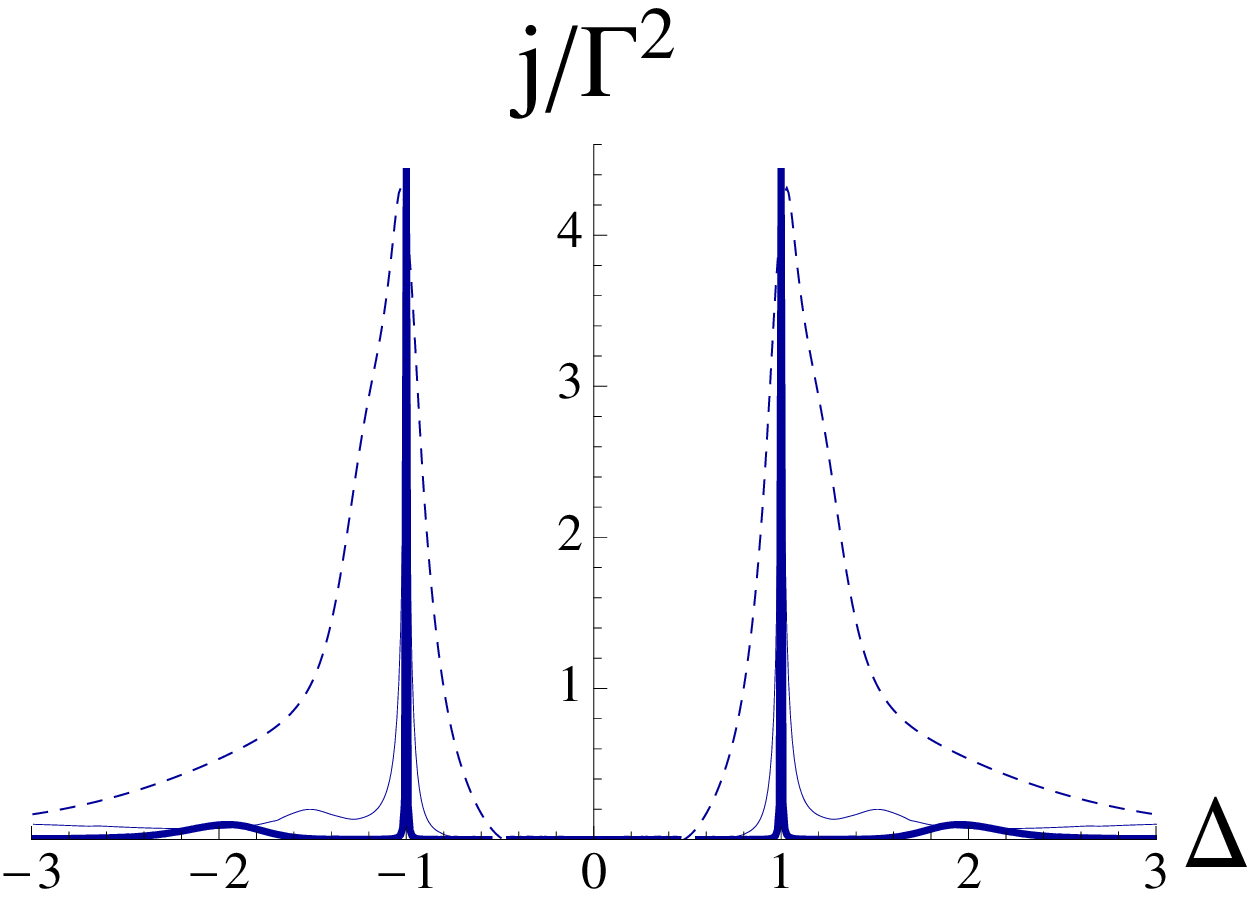}}
\end{center}
\caption{Typical dependence of renormalized current $j/\Gamma^{2}$ versus
anisotropy for small couplings $\Gamma$ and odd system size (Panels (a),(c))
and for even system size (Panels (b),(d)). Dashed, thin, and thick lines
correspond to $\Gamma=10^{-1},10^{-2},10^{-3}$ respectively. The system size
is $N=3,4,5,6$ for Panels (a),(b),(c) and (d), respectively. The curves are
obtained by a numerical solution of the LME. }%
\label{fig:XY}%
\end{figure}

Although analytical expressions of the current are very difficult to obtain
for larger values of $N$ (e.g. $N>4$), we can infer from numerical
calculations that similar behavior exist also in these more complicated cases.
Indeed, from the exact expressions reported above and from direct numerical
calculations it is possible to extrapolate the values of the current for
arbitrary $N$ at the points $\Delta\pm1$ in the small coupling limit as:
\begin{equation}
\lim_{\Gamma\rightarrow0}\lim_{ \Delta\rightarrow\pm1}\left.  \frac{j(\Gamma
)}{\Gamma^{2}}\right\vert _{N} = \frac89 (N-1),\label{generalcurrent}%
\end{equation}
hinting at $\Gamma=0$ becoming a singular point in the thermodynamic limit.
One can readily see that our conjecture (\ref{generalcurrent}) coincides with
the exact values reported in Eqs. (\ref{J2singularity}),
(\ref{N=3singularity0}), (\ref{J4singularity}) for cases $N=2,3,4$,
respectively. Moreover, numerical results provides us a high confidence about
the validity of Eq. (\ref{generalcurrent}) for arbitrary $N$. More precisely,
we find from direct numerical solutions of the LME that the peaks of the
current for $N=7,8,9$ at $\Gamma=10^{-5}$ are: $5.3333, 6.2222, 7.111$,
respectively, in perfect agreement with the prediction of
Eq.(\ref{generalcurrent}). Also note from Fig. \ref{fig:XY} that the numerical
values of the current peaks depicted in panels (c),(d) for cases $N=5,6$
($3.5556$ and $4.4444$, respectively) are in excellent agreement with Eq.
(\ref{generalcurrent}). Moreover, using the Matrix Product Ansatz \cite{MPA},
we were able to check the conjecture (\ref{generalcurrent}) analytically up to
sizes $N=100$ (details to be published elsewhere).  From the physical point of
view, Eq. (\ref{generalcurrent}) is quite interesting because it implies that
in the weak coupling limit an addition of an extra spin to a finite chain
contributes to the total current by a "quantum" of $8\Gamma^{2}/9 $. Since the
spin current is bounded, the Eq.(\ref{generalcurrent}) hints at a singularity
in $\Gamma=0$ point in the thermodynamic limit $N \rightarrow\infty$. Further
details and an analytical proof of (\ref{generalcurrent}) will be presented elsewhere.

\begin{figure}[ptbh]
\begin{center}
\subfigure[\label{fig:JE:aa}]
{\includegraphics[width=0.4\textwidth]{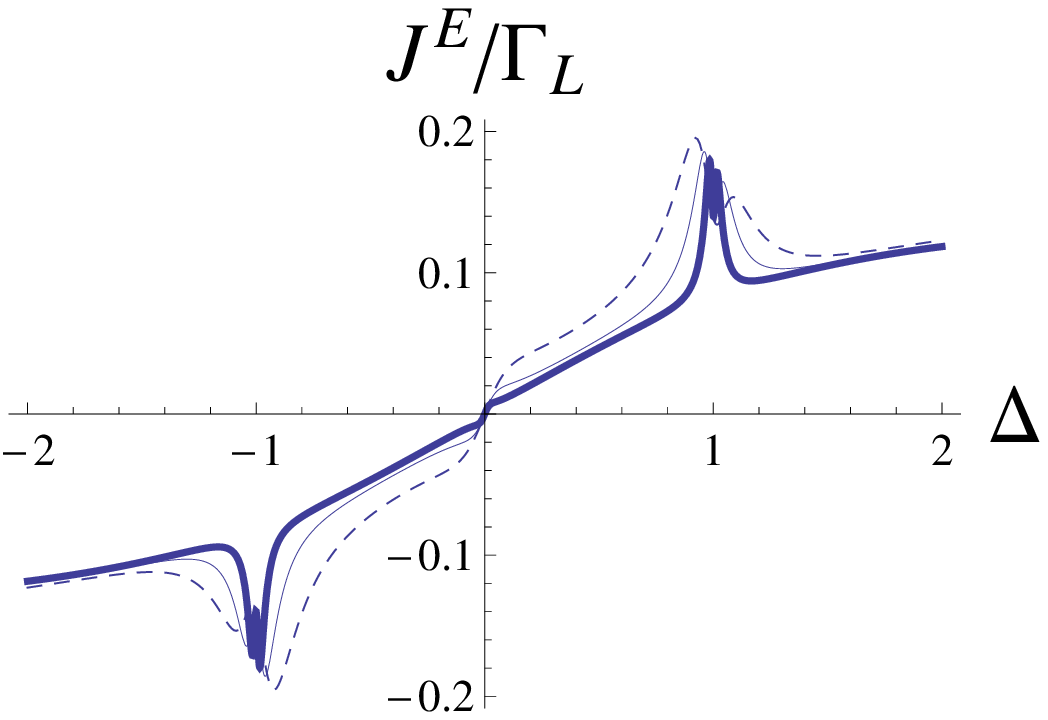}} \qquad
\subfigure[\label{fig:JE:bb}]
{\includegraphics[width=0.4\textwidth]{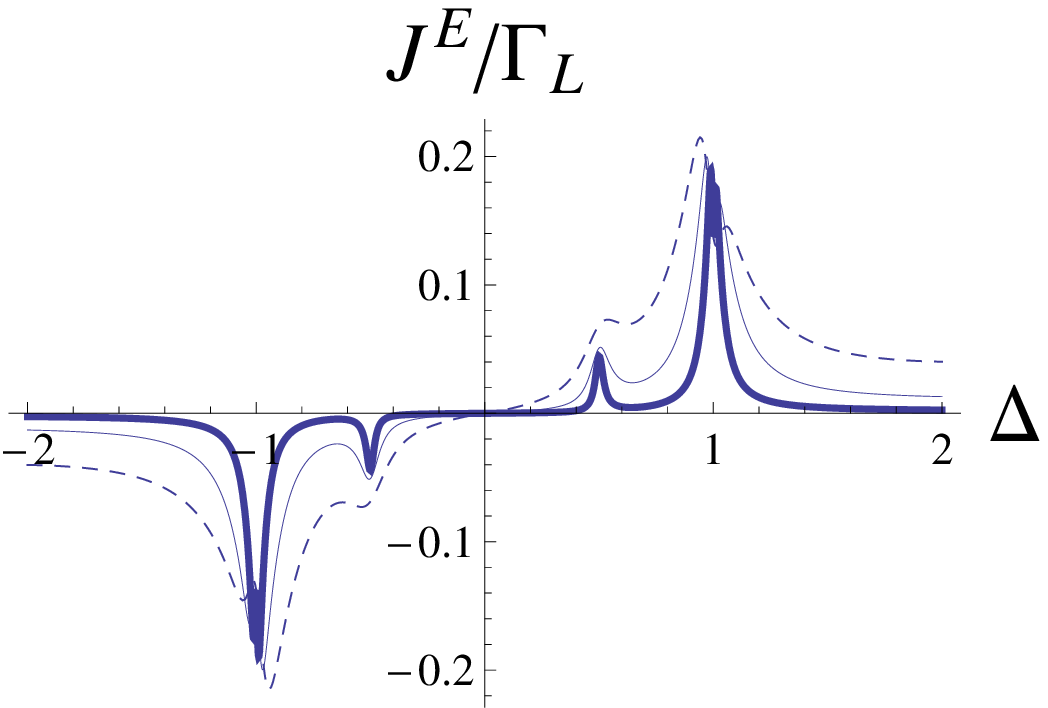}} \qquad
\subfigure[\label{fig:JE:cc}]
{\includegraphics[width=0.4\textwidth]{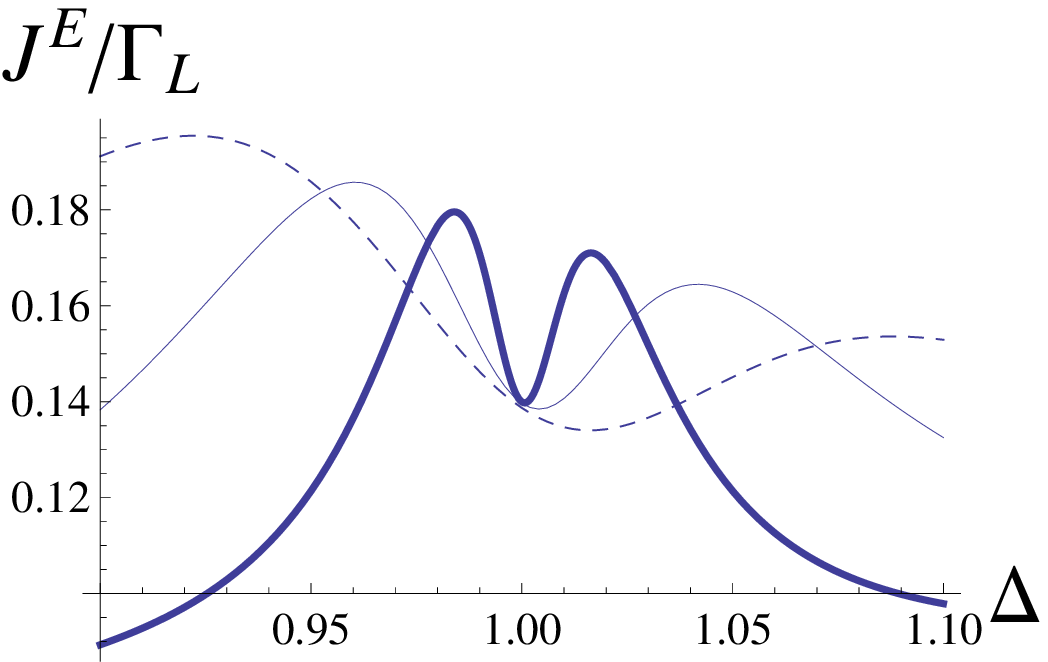}}
\end{center}
\caption{Dependence of renormalized energy current $j/\Gamma_{L}$ versus
anisotropy for small couplings $\Gamma_{L},\Gamma_{R}$ and system size $N=3$
(Panel (a)) and for $N=4$ (Panel (b)). Dashed, thin, and thick lines
correspond to $\Gamma_{L}=0.1,0.05,0.02$ respectively, while $\Gamma
_{R}=\Gamma_{L}/2$. The curves are obtained by numerical solution of the LME.
The peaks for $N=4$ are at the points $\Delta=\pm0.5,\pm1$. Panel (c). A
closeup view of the Panel(a), demonstrating double peak structure close to
$\Delta=1$.}%
\label{fig:JE}%
\end{figure}

One naturally expects that the noticed singular behaviour of the driven $XXZ$
model at the isotropic point is not restricted to the magnetization current
but can be seen for other observables as well. Here we show that the energy
current is also anomalous at the isotropic point. Our purpose here, in
addition, is to demonstrate, that the global symmetry (\ref{SymmetryGlobal})
is not crucial for singularity, because the energy current only flows if the
symmetry (\ref{SymmetryGlobal}) is broken. Indeed, the energy current operator
$\hat{J}_{n}^{E}(\Delta)$, defined through the local Hamiltonian term
$h_{n,n+1}$ as $dh_{n,n+1}/dt=\hat{J}_{n-1}^{E}(\Delta)-\hat{J}_{n}^{E}%
(\Delta),$
\[
\hat{J}_{n}^{E}(\Delta)=-\sigma_{n}^{z}\hat{\jmath}_{n-1,n+1}+\Delta
(\hat{\jmath}_{n-1,n}\sigma_{n+1}^{z}+\sigma_{n-1}^{z}\hat{\jmath}_{n,n+1})
\]
where $\hat{\jmath}_{n,m}=2(\sigma_{n}^{x}\sigma_{m}^{y}-\sigma_{n}^{y}%
\sigma_{m}^{x})$, changes sign under the action of (\ref{SymmetryGlobal}),
$(RU_{rot}^{+}\Sigma_{x})\hat{J}_{n}^{E}(RU_{rot}^{+}\Sigma_{x})$ $=-\hat
{J}_{n}^{E}$ and therefore in the stationary state it is strictly zero
$\langle\hat{J}_{n}^{E}\rangle=J^{E}=0$ (as a local integral of motion, the
energy current does not depend on the choice of the bond $n$). Lifting up the
symmetric choice for the couplings to the left and right reservoirs in
(\ref{LME}), i.e., considering the Lindblad equation (\ref{LME}) with
$\Gamma_{L}\neq\Gamma_{R}$ we break the symmetry (\ref{SymmetryGlobal}) and
the energy current becomes permissible, see also \cite{PopkovLivi}. The
current of energy for small couplings $\Gamma_{L},\Gamma_{R}\ll1,$ with a
fixed ratio $\Gamma_{L}/\Gamma_{R}=const$ shows, again, an approach to a
singularity in the isotropic point, but with a different scaling,
$J^{E}(\Delta=1)=O(\Gamma_{L})$, and qualitatively different behaviour, see
Fig.\ref{fig:JE}. Indeed, the anomaly of energy current at the isotropic point
has a shape of a twin peak, which, as $\Gamma_{L}$ decreases, becomes more and
more narrow, see Fig.\ref{fig:JE:cc}, and for $\Gamma_{L}\rightarrow0$
develops a twin peak singularity. For $N=4$ we see a twin peak singularity at
$\Delta=1$ and a single peak singularity at $\Delta=0.5$, illustrated on
Fig.\ref{fig:JE}(b). Unlike the spin current which is the odd (even) function
of $\Delta$ for odd (even) number of sites, the energy current is always an
odd function of $\Delta$. These parity features are direct consequences of the
mappings (\ref{SymmetryEven}) and (\ref{SymmetryOdd}), see also
\cite{PopkovXYtwist}. Further details will be discussed elsewhere. Here we
just mention another observable which has a singularity at the isotropic
point, also in the symmetric case $\Gamma_{L}=\Gamma_{R}=\Gamma$:\ it is the
actual value of magnetizations at the boundaries: $\langle\sigma_{1}%
^{x}\rangle,\langle\sigma_{N}^{x}\rangle$, $\langle\sigma_{1}^{y}\rangle
$,$\langle\sigma_{N}^{y}\rangle$, which, as $\Gamma\rightarrow0$, remain
finite at $\Delta=1$, and attain different values at the point $\Delta1\pm0$,
data not shown. Interestingly, this singularity cancels in the difference
$\Delta x=\langle\sigma_{1}^{x}-\sigma_{N}^{x}\rangle$ and $\Delta
y=\langle\sigma_{1}^{y}-\sigma_{N}^{y}\rangle$: the actual magnetization
differences $\Delta x,\Delta y$ are regular at $\Delta=1$.

At this point one might suspect, that the singularities in the currents of
spin and energy are just artefacts (or direct consequences) of the singular
behaviour of the boundary gradients. To eliminate the influence of the
irregular behaviour of the boundary gradients, in the next section we consider
another limit, where we control not only the boundary gradients $\Delta
x,\Delta y$, but also the individual magnetizations at the boundaries
$\langle\sigma_{1}^{x}\rangle,\langle\sigma_{1}^{y}\rangle$, $\langle
\sigma_{1}^{z}\rangle$,$\langle\sigma_{N}^{x}\rangle,\langle\sigma_{N}%
^{y}\rangle$, $\langle\sigma_{N}^{z}\rangle$, which become exactly equal to
$0,-\kappa,0,\kappa,0,0$, respectively.

\section{Case of weak boundary gradients and strong coupling}

\label{sec::WeakGradient} 
\begin{figure}[ptbh]
\begin{center}
\subfigure[\label{fig:XYa}]
{\includegraphics[width=0.4\textwidth]{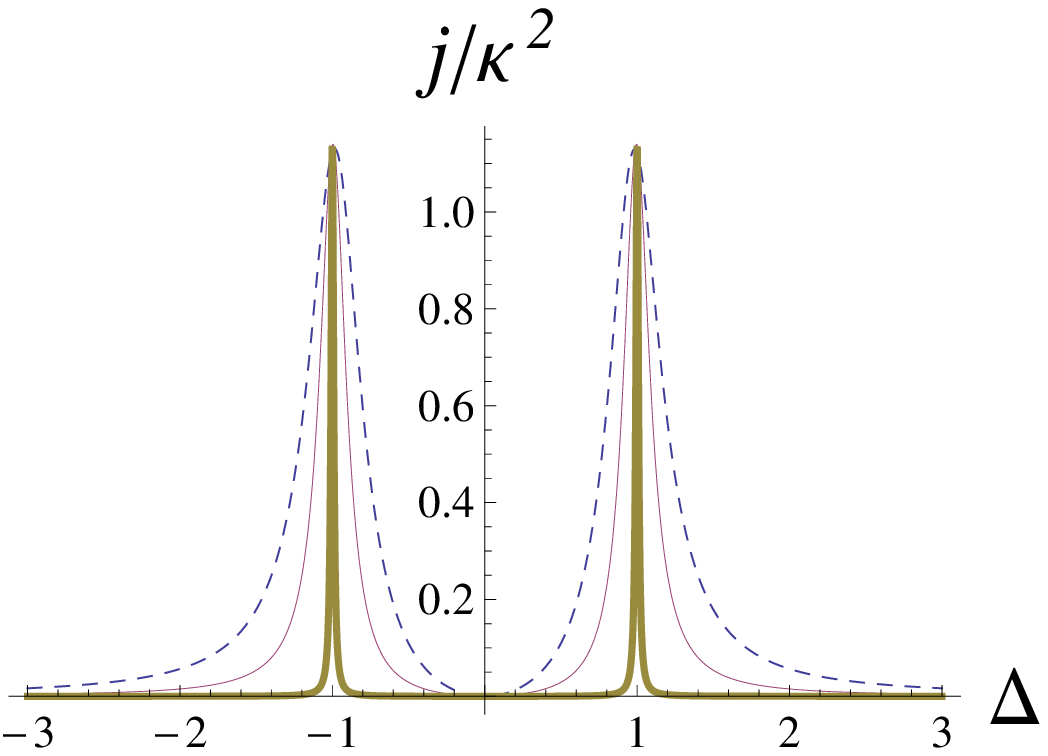}} \qquad
\subfigure[\label{fig:XYb}]
{\includegraphics[width=0.4\textwidth]{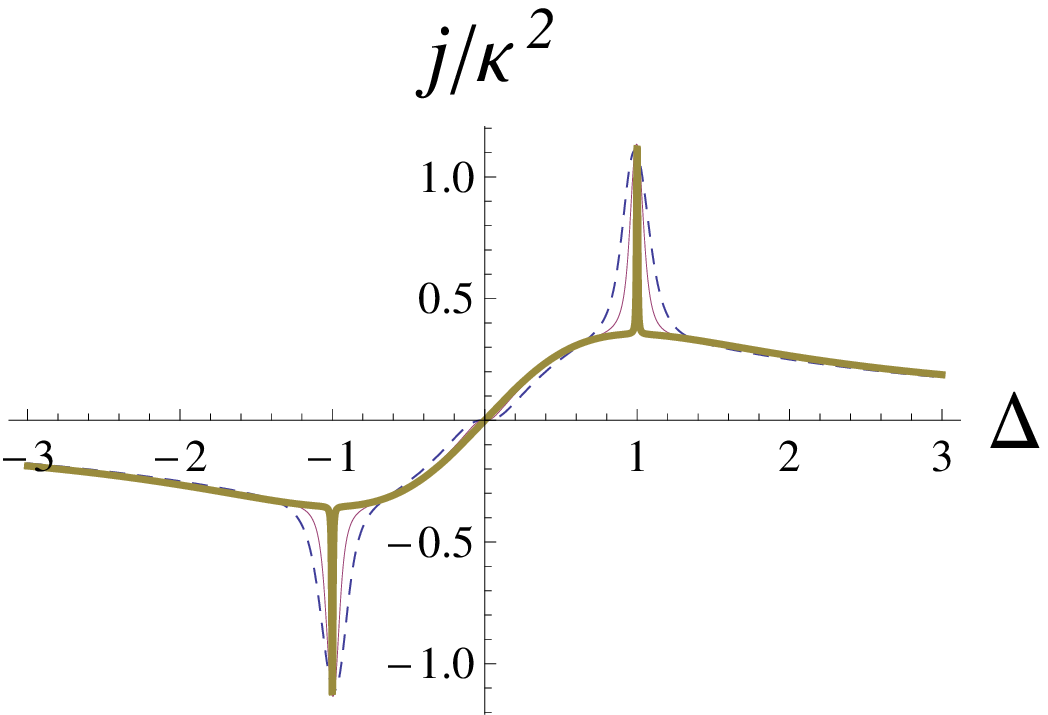}} \qquad
\end{center}
\caption{Typical dependence of renormalized current $j/\kappa^{2}$ versus
anisotropy for small driving $\kappa$ and even system size (Panels (a)) and
for odd system size, Panel (b). Dashed, thin, and thick lines correspond to
$\kappa=0.2,0.1,10^{-2}$ respectively. The system size $N=4$ for Panels (a)
and $N=5$ for Panels (b). The curves are given by exact analytic expressions
(\ref{JJkappaN4}) for $N=4$ and analogous analytic expression (not reported)
for $N=5$.}%
\label{Fig_SmallDriving}%
\end{figure}

In this section we investigate the weak gradient and
strong coupling case ($\kappa\ll1$,$\Gamma^{-1}\ll1$) by means of the
perturbative approach for the LME in powers of $\Gamma^{-1}$, developed in
\cite{PopkovXYtwist}. Note that while in the standard derivation of the
Lindblad Master equation one usually assumes the coupling to the reservoirs to
be weak\cite{Petruccione},\cite{Wichterich07}, in the ancilla Master equation
construction \cite{ClarkPriorMPA2010} this restriction is lifted and the
effective coupling can become arbitrarily strong. We also remark that the
uniqueness of the nonequilibrium steady state for any coupling $\Gamma$ is
guaranteed by the completeness of the algebra, generated by the set of
operators $\{H,V_{m},W_{m},V_{m}^{+},W_{m}^{+}\}$ under multiplication and
addition \cite{EvansUniqueness}, and is verified straightforwardly along the
lines \cite{ProsenUniqueness}. We search for a stationary solution of the
Lindblad equation in the form of a perturbative expansion in $\Gamma^{-1}$,
\begin{equation}
\rho(\Delta,\Gamma)=\sum_{k=0}^{\infty}\left(  \frac{1}{2\Gamma}\right)
^{k}\rho_{k}(\Delta),\label{PT_largeCouplings}%
\end{equation}
where $\rho_{0}(\Delta)=\lim_{\Gamma\rightarrow\infty}\rho(\Delta,\Gamma)$
satisfies $\mathcal{L}_{LR}[\rho_{0}(\Delta)]=0$. Here and below we denote by
$\mathcal{L}_{LR}[.]$ the sum of Lindblad boundary actions $\mathcal{L}%
_{LR}[.]=\mathcal{L}_{L}[.]+\mathcal{L}_{R}[.]$. This enforces a factorized
form
\begin{equation}
\rho_{0}(\Delta,\kappa)=\rho_{L}\otimes\left(  \left(  \frac{I}{2}\right)
^{\otimes_{N-2}}+M_{0}(\Delta,\kappa)\right)  \otimes\rho_{R}%
,\label{InitialChoiceRo0}%
\end{equation}
where $\rho_{L}$ and $\rho_{R}$ are one-site density matrices given by
(\ref{RoL}),(\ref{RoR}) which satisfy $\mathcal{L}_{L}[\rho_{L}]=0$,
$\mathcal{L}_{R}[\rho_{R}]=0$ and $M_{0}(\Delta,\kappa)$ is a matrix to be
determined self-consistently later. Below we shall drop $\Delta$- and
$N$-dependence in $\rho_{k}$ and $M_{k}$ for brevity of notations.
Substituting (\ref{PT_largeCouplings}) into (\ref{LME}), and comparing the
orders of $\Gamma^{-k}$, we obtain recurrence relations%
\begin{equation}
i[H,\rho_{k}]=\frac{1}{2}\mathcal{L}_{LR}\rho_{k+1}\,\,k=0,1,2...
\end{equation}
A formal solution of the above is $\rho_{k+1}=-2\mathcal{L}_{LR}^{-1}%
(Q_{k+1})$ where $Q_{k+1}=-i[H,\rho_{k}]$. Note however that the operator
$\mathcal{L}_{LR}$ has a nonempty kernel subspace, and is not invertible on
the elements from it. The kernel subspace $\ker(\mathcal{L}_{LR})$ consists of
all matrices of type $\rho_{L}\otimes A\otimes\rho_{R}$ where $A$ is an
arbitrary $2^{N-2}\times$ $2^{N-2}$ matrix. Therefore $\rho_{k+1}$ exists only
if $[H,\rho_{k}]\cap\ker(\mathcal{L}_{LR})=\varnothing$, which in our case
reduces to a requirement of a null partial trace, see \cite{PopkovXYtwist}.
\begin{equation}
Tr_{1,N}([H,\rho_{k}])=0,\,\,k=0,1,2....\label{SecularConditions}%
\end{equation}
which we call secular conditions. Finally, $\rho_{k+1}$ is defined up to an
arbitrary element from $\ker(\mathcal{L}_{LR})$, so we have
\begin{equation}
\rho_{k+1}=2\mathcal{L}_{LR}^{-1}(i[H,\rho_{k}])+\rho_{L}\otimes
M_{k+1}\otimes\rho_{R},\;\;k=0,1,2...\label{Recurrence}%
\end{equation}
Eqs. (\ref{InitialChoiceRo0}), (\ref{SecularConditions}) and (\ref{Recurrence}%
) define a perturbation theory for the Lindblad equation (\ref{LME}) for
strong couplings. At each order of the perturbation theory the secular
conditions (\ref{SecularConditions}) must be satisfied, which impose
constraints on $M_{k}$. Our aim here is to determine the exact NESS in the
limit $\Gamma\rightarrow\infty$ for which considering the two first orders
$k=0,1$ has proved to be enough. The case $N=2$ is trivial since $\rho
_{k}=\rho_{L}\otimes\rho_{R}$. For $N=3$, the most general form of the
matrices $M_{0}(\Delta)$, $M_{1}(\Delta)$, by virtue of the symmetry
(\ref{SymmetryGlobal}), is $M_{0}=b_{0}(\sigma^{x}-\sigma^{y})$, $M_{1}%
=b_{1}(\sigma^{x}-\sigma^{y})$ where $b_{0},b_{1}$ are unknown constants. The
secular conditions (\ref{SecularConditions}) for $k=0$ do not give any
constraints on $b_{0}$, while those for $k=1$ give two nontrivial relations
$b_{1}=0$ and $-4\Delta\kappa^{2}+(3+4\Delta^{2}-\kappa^{2})b_{0}=0$ from
which both $M_{0}$ and $M_{1}$ are determined. For $N=4$ each of the matrices
$M_{0}(\Delta),M_{1}(\Delta)$ compatible with the symmetry
(\ref{SymmetryGlobal}), contains $9$ unknowns, all fixed by the secular
conditions (\ref{SecularConditions}) for $k=0,1$, and so on. For $N=3$ we
obtain
\begin{equation}
\left.  M_{0}(\Delta)\right\vert _{N=3}=\frac{4\Delta\kappa^{2}}{3+4\Delta
^{2}-\kappa^{2}}(\sigma^{x}-\sigma^{y}).\label{M0_N3}%
\end{equation}
The respectively stationary current $j_{0}(\kappa)=\lim_{\Gamma\rightarrow
\infty}j(\kappa,\Gamma)$
\begin{equation}
\left.  j_{0}(\kappa)\right\vert _{N=3}=\frac{8\Delta\kappa^{2}}{3+4\Delta
^{2}-\kappa^{2}}\label{JJkappaN3}%
\end{equation}
does not have any non-analyticity as $\kappa\rightarrow0$, unlike in the case
of weak coupling. However, for $N\geq3$ the singularity sets in. For $N=4$ the
exact expression for the current is
\begin{equation}
\left.  j_{0}(\kappa)\right\vert _{N=4}=\frac{8\Delta^{2}\kappa^{4}\left(
16\Delta^{4}+2\Delta^{2}\left(  5\kappa^{4}-11\kappa^{2}+24\right)
-8\kappa^{6}+25\kappa^{4}-36\kappa^{2}+27\right)  }{D(\Delta,\kappa
)},\label{JJkappaN4}%
\end{equation}
where
\begin{eqnarray*}
& D(\Delta,\kappa)=16\Delta^{8}\left(  \kappa^{2}+3\right)
+4\Delta ^{6}\left(  \kappa^{6}+\kappa^{4}+48\kappa^{2}-6\right)
+\Delta^{4}\left(
-6\kappa^{8}+21\kappa^{6}-\right.  \nonumber\\
& \left.  133\kappa^{4}+339\kappa^{2}-69\right)  \Delta^{2}\left(  9\kappa
^{8}-25\kappa^{6}-7\kappa^{4}+117\kappa^{2}+18\right)  \nonumber\\
& -2\kappa^{8}-\kappa^{6}+3\kappa^{4}-27\kappa^{2}+27,
\end{eqnarray*}
and it does contain a singularity at $\Delta=1$. Indeed, for $N=4$
we observe a non-commutativity of the limits $\Delta\rightarrow1$
and $\kappa \rightarrow0$, signalizing the presence of a
singularity,
\begin{equation}
\lim_{\Delta\rightarrow1}\lim_{\kappa\rightarrow0}\left.  \frac{j_{0}(\kappa
)}{\kappa^{2}}\right\vert _{N=4}=0\label{N=4singularity0}%
\end{equation}%
\begin{equation}
\lim_{\kappa\rightarrow0}\lim_{\Delta\rightarrow1}\left.  \frac{j_{0}(\kappa
)}{\kappa^{2}}\right\vert _{N=4}=\frac{8}{7},\label{N=4singularity1}%
\end{equation}
easily verifiable from (\ref{JJkappaN4}). Indeed, taking the $\lim
_{\kappa\rightarrow0}$ first, we expand $j(\kappa)$ in orders of $\kappa$ and
find the first nonzero contribution at the fourth order,
\begin{equation}
\lim_{\kappa\rightarrow0}\left.  \frac{j_{0}(\kappa)}{\kappa^{4}}\right\vert
_{N=4}=\frac{8\Delta^{2}\left(  4\Delta^{2}+9\right)  }{3\left(  \Delta
^{2}-1\right)  ^{2}\left(  4\Delta^{2}+3\right)  },
\end{equation}
which is singular at $\Delta=1$. As a consequence, the current at the point
$\Delta=1$ (and, in virtue of the symmetry (\ref{SymmetryEven}), also
$\Delta=-1$) has a different scaling ($j_{0}(\kappa,\Delta=\pm1)\sim\kappa
^{2}$) then all other points $\Delta\neq\pm1$, where the current scales as
$\kappa^{4}$, see Fig.\ref{Fig_SmallDriving}. Note that the singularity type
for weak driving $\kappa$ is exactly the same as the one described in
Sec.\ref{sec::WeakCoupling} for even number of sites and weak coupling
$\Gamma$.

For odd-sized system $N=5$ the exact expression for the current is very
complicated and some limiting cases are reported in the
\ref{app::Current for N=5}. Analyzing the analytic expression in the limit
$\kappa\ll1 $ we find the scaling of the type $j^{(0)}_{N=5}(\kappa
)=\alpha(\Delta) \kappa^{2}$ where the prefactor is $\alpha(\Delta)$ is
singular at $\Delta=1$, namely $\alpha(1) \rightarrow8/7$ and $\alpha(1\pm0)
\rightarrow64/181$ as $\kappa\rightarrow0$, see also
Fig.\ref{Fig_SmallDriving}. This type of singularity is exactly the same as
the one seen in Sec.\ref{sec::WeakCoupling} for vanishing $\Gamma$ and odd
system sizes.

As the system size increases, the order of polynomials grow with system size
and becomes complicated. However, at the qualitative level we see the same
behaviour as discussed above for $N=4$ ( $N=5$ ) for system of even (odd) sizes.

The qualitative similarity in the magnetization current behaviour in the cases
(i) of weak coupling and large driving, and (ii) weak driving and large
coupling is not a trivial one. In the case (ii), the amplitude of the
effective $x$- and $y$- boundary gradients is exactly equal to $\kappa$,
independently on the anisotropy $\Delta$, so in the limit $\kappa\ll1$ it
becomes infinitesimally small. Also the individual boundary magnetizations
$\langle\sigma_{1}^{\alpha}\rangle$,$\langle\sigma_{N}^{\alpha}\rangle
$,$\alpha=x,y,z$ are bounded by $\kappa$.

In the case of weak coupling $\Gamma\ll1$, the amplitude of the boundary
gradient at the isotropic point also vanishes in the limit $\Gamma
\rightarrow0$, but the individual boundary magnetizations $\langle\sigma
_{1}^{\alpha}\rangle$,$\langle\sigma_{N}^{\alpha}\rangle$ for $\alpha=x,y$
remain finite and are discontinuous at $\Delta=1$.

The presence of the singularity and non-commutativity of the limits
$\lim_{\Delta\rightarrow1}$ and $\lim_{\Gamma\rightarrow0}$(or $\lim
_{\kappa\rightarrow0}$) must be related to the existence of an additional
symmetry which the $XXZ$ model acquires at the isotropic point $\Delta=1$.
This symmetry however is not the full rotational symmetry of a unitary quantum
$XXX$ Hamiltonian, since the dissipative Lindblad terms in LME do not have the
full rotational symmetry. On the other hand, qualitative differences between
singularities for odd and even system sizes $N$, seen for spin current, energy
current, and other observables, is a consequence of intrinsic properties of
the XY-twisted model, which is manifested by (\ref{SymmetryEven}%
),(\ref{SymmetryOdd}).

\section{Conclusions}

\label{sec::Conclusions}

We have investigated an $XXZ$ spin chain coupled at the ends to a dissipative
boundary reservoirs, which impose a twisting angle between the first and the
last spin in the $XY$ plane, described by a Lindblad Master equation. We
pointed out the non-analytic character of the non-equilibrium steady state in
the two limits: of vanishing coupling $\Gamma$ and of vanishing driving
$\kappa$. Unlike in the approaches proposed before, where a small but fixed
boundary driving (of the order of $10^{-2}$) is being typically used, here we
access arbitrary small coupling or driving analytically. Our approach allows
to establish a presence of a singularity in the NESS for $\Gamma\rightarrow0$
and $\kappa\rightarrow0 $, which is expected to be present in the system for
all system sizes. The singularity is evidenced on an example of several
observables: the magnetization current, the energy current, and the boundary
magnetizations. For magnetization current, the analytic treatment is presented.

The character of the singularity qualitatively depends on a parity of system
size $N$. For odd $N$ we find that the spin current scales for small $\Gamma$
or small $\kappa$ as $j=f(\Delta)\Gamma^{2},j=g(\Delta)\kappa^{2}$, where for
vanishing $\Gamma$ or $\kappa$ the functions $f$ and $g$ have different finite
values at $\Delta=1$, and $\Delta=1\pm0$. For even sizes $N$, the spin current
still scales quadratically with $\Gamma$ or $\kappa$ at the point $\Delta=1$,
while it scales as $j=f_{1}(\Delta)\Gamma^{4},j=g_{1}(\Delta)\kappa^{4}$ at
all other points. The energy current $J^{E}$ scales in the isotropic point
linearly with couplings $\Gamma_{L},\Gamma_{R}$ (which must be different,
$\Gamma_{L}\neq\Gamma_{R}$), and develops a twin peak singularity. Here we are
not already in a position to discuss the conductivity of spin or energy due to
the small system sizes; however the DMRG studies
\cite{ZnidaricJStat2010_2siteLindblad} and recently proposed exact approaches
might give access to large system sizes and even to the thermodynamic limit.

\textbf{Acknowledgements} VP acknowledges the Dipartimento di Fisica e
Astronomia, Universit\`a di Firenze, for support through a FIRB initiative.
M.S. acknowledges support from the Ministero dell' Istruzione, dell'
Universit\'a e della Ricerca (MIUR) through a \textit{Programma di Ricerca
Scientifica di Rilevante Interesse Nazionale} (PRIN)-2010 initiative.

\appendix

\section{Current for $N=5$ (weak driving limit)}

\label{app::Current for N=5}

For $N=5$ and in the limit $\Gamma\rightarrow\infty$ the exact expression for
the current is given by a complicated expression. We report different limits
here below. Taking the limit $\lim_{\kappa\rightarrow0}$ first, we find%

\begin{equation}
\lim_{\kappa\rightarrow0}\frac{j_{0}(\kappa,\Delta)}{\kappa^{2}}%
=\frac{16\Delta\left(  \Delta^{2}+3\right)  }{28\Delta^{4}+81\Delta^{2}+72},
\end{equation}
and consequently $\lim_{\Delta\rightarrow1}\lim_{\kappa\rightarrow0}%
\frac{j_{0}(\kappa,\Delta)}{\kappa^{2}}=\frac{64}{181}$. On the other hand,
taking the limit $\lim_{\Delta\rightarrow1}$ first, we find $\lim
_{\Delta\rightarrow1}\frac{j_{0}(\kappa,\Delta)}{\kappa^{2}}=-\frac{F}{G}$, where%

\begin{eqnarray*}
&  F=8(128\kappa^{22}-2128\kappa^{20}+5792\kappa^{18}+44096\kappa
^{16}-245827\kappa^{14}+182930\kappa^{12}+\nonumber\\
&  980805\kappa^{10}-1523750\kappa^{8}-1515697\kappa^{6}+3477286\kappa
^{4}-1171473\kappa^{2}+2235870),\nonumber
\end{eqnarray*}
\begin{eqnarray*}
&  G=576\kappa^{22}+856\kappa^{20}-34736\kappa^{18}-102787\kappa
^{16}+1669223\kappa^{14}-4033065\kappa^{12}-\nonumber\\
&  2004745\kappa^{10}+13257113\kappa^{8}+3528813\kappa^{6}-16075891\kappa
^{4}-7450779\kappa^{2}-15651090\nonumber
\end{eqnarray*}

and consequently $\lim_{\kappa\rightarrow0}\lim_{\Delta\rightarrow1}%
\frac{j_{0}(\kappa,\Delta)}{\kappa^{2}}=\frac{8}{7}$. Another way to see the
discontinuity is to estimate the derivatives $\left.  (\partial^{m}%
j_{0}(\kappa,\Delta)/\partial\kappa^{m})\right\vert _{\kappa=0}$, all even
orders of which, starting from $m=4$, have singularities at $\Delta=1$ as
\[
\left.  \frac{\partial^{2n}}{\partial\kappa^{2n}}j_{0}(\kappa,\Delta
)\right\vert _{\kappa=0}=O\left(  \frac{1}{(\Delta-1)^{2n-2}}\right)  .
\]
For odd $m=2n+1$, we find $\left.  (\partial^{2n+1}j_{0}(\kappa,\Delta
)/\partial\kappa^{2n+1})\right\vert _{\kappa=0}=0$.


\section*{References}

\begin{thebibliography}{99}                                                                                               %
\bibitem {Petruccione}H.-P. Breuer and F. Petruccione, \textit{The Theory of
Open Quantum Systems}, Oxford University Press, (2002).

\bibitem {PlenioJumps}M.B. Plenio and P.L Knight, Rev. Mod. Phys. \textbf{70},
101 (1998).

\bibitem {Wichterich07}H. Wichterich, M. J. Henrich, H.P. Breuer, J. Gemmer
and M. Michel, \textit{Phys.Rev. E} \textbf{76} ,\ 031115 (2007)

\bibitem {reviewBrenig07}F. Heidrich-Meisner, A. Honecker, and W. Brenig,
\textit{Eur. Phys. J. Special Topics} \textbf{151}, 135 (2007), and references therein.

\bibitem {zotos}X. Zotos, \textit{J. Phys. Soc. Jpn. Supp.} \textbf{74}, 173
(2005) and references therein.

\bibitem {KlumperLectNotes2004}A. Klumper, Lect. Notes Phys. \textbf{645}, 349 (2004).

\bibitem {Bosonisation}F. M. D. Haldane, \textit{J. Phys. C} \textbf{14},
2585(1981); A. O. Gogolin and N. V. Prokof'ev, \textit{Phys. Rev. B}
\textbf{50}, 4921 (1994); Schulz H, 1996, in: \textquotedblleft Correlated
Fermions and Transport in Mesoscopic Systems\textquotedblright, Les Arcs,
Savoie, (Editions Frontieres, 1996)

\bibitem {SchollwoeckRev05}U Schollw\"{o}ck, R\textit{ev. Mod. Phys.}
\textbf{77}, 259 (2005).

\bibitem {HessBallistic2010}N. Hlubek, P. Ribeiro, R. Saint-Martin, A.
Revcolevschi, G. Roth, G. Behr, B. B\"{u}chner, and C. Hess, P\textit{hys.
Rev. B} \textbf{81}, 020405 (2010).

\bibitem {TP_Pizorn08}T. Prosen and I. Pizorn, \textit{Phys. Rev. Lett.}
\textbf{101}, 105701 (2008).

\bibitem {BenentiPRB2009}G. Benenti, G. Casati, T. Prosen, D. Rossini and M.
\v{Z}nidari\v{c}, \textit{Phys. Rev. B} \textbf{80}, 035110 (2009).

\bibitem {ZunkovichJStat2010ExactXY}B. \v{Z}unkovi\v{c} and T. Prosen,
\textit{J. of Stat. Mech.} P08016 (2010).

\bibitem {Pros08}T. Prosen, \textit{New. J. Phys.} \textbf{10}, 043026 (2008).

\bibitem {ProsenExact2011}T. Prosen, \textit{Phys. Rev. Lett.} \textbf{107},
137201 (2011).

\bibitem {MPA}D. Karevski, V. Popkov and G. Sch\"{u}tz, \textit{Phys. Rev.
Lett. }\textbf{110}, 047201 (2013)













\bibitem {znidaricprl2011}M. \v{Z}nidari\v{c},\textit{ Phys. Rev.Lett.}
\textbf{106}, 220601 (2011).

\bibitem {KuboBook}R. Kubo , M. Toda, N. Hashitsume, Statistical Physics II:
Nonequilibrium Statistical Mechanics (Springer Series in Solid-State Sciences)

\bibitem {Lindblad2011}V. Popkov, Mario Salerno and G. M. Sch\"{u}tz,
\textit{Phys. Rev. E }\textbf{85}, 031137 (2012).

\bibitem {PopkovXYtwist}V. Popkov, \textit{J. Stat. Mech.} (2012) P12015

\bibitem {PopkovLivi}V. Popkov and R. Livi, \textit{New J. Phys.} \textbf{15}
(2013) 023030

\bibitem {ClarkPriorMPA2010}S. R. Clark, J. Prior, M. J. Hartmann, D. Jaksch
and M. B. Plenio, New J. of Phys. \textbf{12}, 025005(2010).

\bibitem {SP2012}M. Salerno and V. Popkov, \textit{Phys. Rev. E} \textbf{87},
022108 (2013)



















\bibitem {BoundaryRelaxationTimes}The relaxation times depend on spin
component. At the left boundary, $x$- and $z$-spin components of $\rho$ relax
as $e^{-4\Gamma t}$ while the $z$-spin component relax as $e^{-2\Gamma t}$,
see \cite{Lindblad2011}

\bibitem {TwistingRemark}If $\kappa<1$, the situation is more subtle since
then the target states $\rho_{L}$ and $\rho_{R}$ are mixed

\bibitem {energyRemark}For symmetric choice of the couplings
(\ref{SymmetryU-Uprime})the energy current disappears as a consequence of the
symmetry (\ref{SymmetryGlobal}), see \cite{PopkovXYtwist}.

\bibitem {EvansUniqueness}D.E. Evans, Comm. Math. Phys, \textbf{54}, 293 (1977)

\bibitem {ProsenUniqueness}T. Prosen, Phys. Scr. \textbf{86}, 058511 (2012).

\bibitem {ZnidaricJStat2010_2siteLindblad}T. Prosen and M. \v{Z}nidari\v{c},
J. Stat. Mech. (2009) P02035; M. \v{Z}nidari\v{c}, J. of Stat. Mech. P12008 (2011).
\end{thebibliography}
\end{document}